# Mechanical properties of the magnetocaloric intermetallic LaFe$_{11.2}$Si$_{1.8}$ alloy at different length scales

Oleksandr Glushko[1], Alexander Funk[2], Verena Maier-Kiener[3], Philipp Kraker[1], Maria Krautz[2], Jürgen Eckert[1,4], Anja Waske[2,5]

[1]*Erich Schmid Institute of Materials Science, Austrian Academy of Sciences, Jahnstrasse 12, A-8700 Leoben, Austria*

[2]*Institute for Complex Materials, IFW Dresden, Helmholtzstraße 20, D-01069 Dresden, Germany*

[3]*Department Physical Metallurgy and Materials Testing, Montanuniversität Leoben, Roseggerstraße 12, A-8700 Leoben, Austria*

[4]*Department Materials Physics, Montanuniversität Leoben, Jahnstraße 12, A-8700 Leoben, Austria*

[5]*Federal Institute for Materials Research and Testing, Unter den Eichen 87, D-12205 Berlin, Germany*

**Abstract**

In this work the global and local mechanical properties of the magnetocaloric intermetallic LaFe$_{11.2}$Si$_{1.8}$ alloy are investigated by a combination of different testing and characterization techniques in order to shed light on the partly contradictory data in recent literature. Macroscale compression tests were performed to illuminate the global fracture behavior and evaluate it statistically. LaFe$_{11.2}$Si$_{1.8}$ demonstrates a brittle behavior with fracture strains below 0.6 % and widely distributed fracture stresses of 180-620 MPa leading to a Weibull modulus of $m$ = 2 to 6. The local mechanical properties, such as hardness and Young's modulus, of the main and secondary phases are examined by nanoindentation and Vickers microhardness tests. An intrinsic strength of the main magnetocaloric phase of at least 2 GPa is estimated. The significantly lower values obtained by compression tests are attributed to the detrimental effect of pores, microcracks, and secondary phases. Microscopic examination of indentation-induced cracks reveals that ductile α-Fe precipitates act as crack arrestors whereas pre-existing cracks at La-rich precipitates provide numerous 'weak links' for the initiation of catastrophic fracture. The presented systematic study extends the understanding of the mechanical reliability of La(Fe, Si)$_{13}$ alloys by revealing the correlations between the mechanical behavior of macroscopic multi-phase samples and the local mechanical properties of the single phases.

# 1. Introduction

Magnetocaloric refrigeration is a technique that has a high theoretical energy efficiency compared to conventional cooling methods based on gas compression and expansion [1, 2]. Cooling of food, medicine, technical processes or buildings as key technology of modern society is a huge business sector and enormous energy consumer market. Magnetocaloric refrigeration has the potential to save energy in cooling applications leading to environmental friendly solutions. The magnetocaloric effect (MCE) on which magnetocaloric refrigeration is based was discovered first in nickel by Weiss and Picard in 1917 [3]. By magnetizing and demagnetizing a magnetocaloric material at its magnetic transition temperature $T_t$, the material will release or consume heat from its surroundings, which can be utilized to build up a heat pump for refrigeration applications. Since the discovery of the giant MCE by Pecharsky et al. [4] in $Gd_5Si_2Ge_2$ with a first order magneto-volume transition, other first and second order MCE alloy classes, e.g. $Fe_2P$-based, Heusler alloys or $La(Fe, Si)_{13}$, are in the scope of research to lower the material costs and to optimize the MCE properties [5]. However, these alloy classes of interest are mechanically brittle and shaping them into an effective heat exchanger or ensuring the necessary cycling stability is challenging.

The intermetallic phase $La(Fe, Si)_{13}$ exhibits a large MCE at its magneto-volume transition temperature $T_t$ from para- to ferromagnetic state, which is accompanied by a change of volume and temperature of the material up to $\Delta T_{ad}$ = 5-7 K in a field change of $\mu_0 H$ = 1.93 T [6]. This enables the possibility to apply this material in a magnetocaloric refrigeration device. Depending on additional elements, e.g. Co or Mn as well as hydrogenation, the transition temperature $T_t$ can be tuned towards room temperature, which is essential for its application as a cooling device at ambient temperature. By doping, the character of the phase transition may change from initially first order in $LaFe_{11.8}Si_{1.2}$ with low silicon content ($T_t$ = 180 K), to second order by adding Co ($T_t$ at room temperature) or a high silicon content $LaFe_{11.2}Si_{1.8}$ ($T_t$ = 220 K). The base $La(Fe, Si)_{13}$ alloy system shows an overall mechanically brittle behavior and a large difference in volume $\Delta V$ of the para- and ferromagnetic phases of up to 1.5 vol% with decreasing Si content [7]. This volume change induces cracks and mechanical degradation of the material over several induced MCE cycles [6, 8, 9, 10], especially in $La(Fe, Si)_{13}$ alloys with a first order phase transition.

A magnetocaloric heat exchanger – the centerpiece of a refrigeration device – must be shaped out of MCE material, for instance as packed bed of particles, arranged plates or rods, drilled holes in a solid body [11] or more complex geometries [12]. To overcome the issue of shaping brittle MCE alloys and to address a high cycling stability, embedding the MCE material into a polymer [13, 14], metallic matrix [15, 16] or metallic cover [17] was suggested. Apart from such advanced approaches for improving the shapeability, knowledge about the basic mechanical properties such as Young's modulus, strength, or hardness of $La(Fe, Si)_{13}$-type alloys is still rather vague. It has been discussed, that the temperature-dependence of the elastic modulus shows a dip at the magnetic transition temperature of $La(Fe, Si)_{13}$-type alloys [18]. The simultaneous volume change gives reason to this behavior. Other alloys, e.g. FeRh [19] or $BaTiO_3$ [20], with temperature dependent structural changes show a characteristic change of the Young's modulus at the transition temperature. However, the mechanical properties at room temperature are important for the machining, shaping and handling of the materials, as machining is usually done at ambient conditions. The results of mechanical testing published so far on $La(Fe, Si)_{13}$-type alloys are divergent or even controversial. For instance, a Young's modulus of about 130 GPa was measured for $La(Fe, Co, Si)_{13}$ alloys by means of nanoindentation in [18]. The Young's modulus estimated from compression tests on similar $La(Fe, Co, Si)_{13}$ material in [21] has a value of about 300 GPa while the fracture strain was only about 0.2 %. Other compressive tests showed completely different results such as fracture strains of more than 5 % but stresses of only 120 MPa with a Young's modulus of only few GPa if estimated from the presented stress-strain curves [22]. Such values most probably stem from the machine stiffness effect which was not taken into account. To evaluate the mechanical behavior of magnetocaloric alloys, knowledge of the mechanical properties is indispensable and may give opportunities to improve the cycling stability of such materials. This is why we followed a systematic approach on different length scales.

The global mechanical properties and macroscopic fracture of homogenized $LaFe_{11.2}Si_{1.8}$ with a typical microstructure including the magnetocaloric main phase and secondary phases are examined by means of compression testing and statistically discussed in terms of a Weibull distribution. The ternary alloy was chosen in order to access the properties of the main magnetocaloric phase (often called "1:13 phase") at ambient conditions. Magnetocaloric transition of this alloy is well below the room temperature so the alloy is in paramagnetic state meaning that the influence of the lattice softening occurring in the ferromagnetic phase during transition on the elastic constants can be neglected. Moreover, the homogeneous grain size distribution and low amount of minority phases compared to alloys with lower silicon content was the reason for choosing a material with a Si content of $x = 1.8$ in $LaFe_{13-x}Si_x$ [23]. The local mechanical properties of the main phase and the secondary phases are described using nanoindentation. Furthermore, the correlations between the propagation of surface cracks and the local microstructure are discussed. Presented results are assumed to be valid for all $LaFe_{13-x}Si_x$ -based alloys with similar microstructure.

## 2. Experimental

### 2.1 Material fabrication and sample preparation

Material buttons of the nominal composition $LaFe_{11.2}Si_{1.8}$ were produced by arc melting of the pure elements La, Fe and Si (99.9% purity). An annealing treatment of the buttons at $T = 1050$ K for $t = 168$ h and subsequent quenching in water was applied. For this, the samples were wrapped in tantalum foil and sealed in quartz tubes with $p = 200$ mbar of argon. For compression tests rectangular bars with dimensions 2x2x10 mm³ were cut out of the buttons by wire erosion. Afterwards the samples were grinded with SiC paper P1200 to achieve plane-parallel surfaces and remove the oxidation layer. These bars were then cut using a diamond wire saw either in two or in four samples with geometries of approximately 2x2x5 mm³ and 2x2x2.5 mm³, respectively. Thus, for mechanical testing two types of samples with the same cross-section (2x2 mm²) but with two different aspect ratios (1:1.2 and 1:2.5) were utilized to investigate the influence of the sample volume on the mechanical properties. For scanning electron microscopy analysis (SEM), X-ray diffraction (XRD) and nanoindentation experiments, several samples were embedded in conductive epoxy, followed by grinding and polishing.

### 2.2 Magnetic measurements

The magnetic properties of the MCE material were evaluated by utilizing a Quantum Design® PPMS-VSM magnetometer. For the temperature dependent magnetization measurements, a temperature sweep rate of $\dot{T} = 0.1$ Kmin$^{-1}$ and a magnetic field of about $\mu_0 H = 10$ mT were applied. By measuring isothermal magnetization curves $M(H)$ in a magnetic field change of $\Delta\mu_0H$ = 2 T between $T$ = 200 K to $T$ = 230 K with a step-size of 1 K, the magnetic entropy change $\Delta S_{mag}$ was calculated by applying the Maxwell relation [11, 24]:

$$\Delta S_{mag} = \int_0^H \frac{\partial M}{\partial T} dH \quad . \quad (1)$$

## 2.3 Microstructural characterization methods

A Zeiss LEO 1525 scanning electron microscope (SEM) was used for examination of the surface and detecting the secondary phases using a backscattered electron (BSE) detector. Energy dispersive X-ray spectroscopy (EDX) was employed to describe the chemical compositions of different phases. Furthermore, electron backscattered diffraction (EBSD) imaging was performed for a detailed analysis of the grain structure. Focused ion beam (FIB) milling was utilized for the cross-sectioning of cracks.

## 2.4 Mechanical properties testing methods

Several different testing methods were employed for the characterization of the mechanical properties of $LaFe_{11.2}Si_{1.8}$ at room temperature. To measure the *global mechanical properties*, such as critical fracture stress and strain, compression tests were performed on an Instron Hydropulse testing device. In order to account for the stiffness of the testing machine, the strain during compression testing was measured directly on the sample surface using a home-made system described as follows. A commercial Canon 750d photo camera equipped with a macro-lens was used to take pictures of the sample during the compression test. The resolution of the recorded images (pixel-to-pixel distance) was about 3.7 µm. The pictures were then analyzed using the freely available digital image correlation (DIC) software GOM Correlate [25]. The DIC software enables measurements of the relative displacements between defined points on the sample surface [26]. For this, a black-white pattern was sprayed on the sample surface to create a proper image contrast for DIC in order to place virtual (digital) extensometers. Each extensometer measures the distance between the marked surface points on each image taken during the deformation. After correlating the strain measured by the extensometers with the stress measured by the force transducer of the tensile testing device, one can obtain the stress-strain curves. The compression tests were performed in displacement-control mode in order to prevent possible damage of the force transducer. A minimum of 15 samples of each sample geometry (1:1.2 and 1:2.5) was tested.

*Local mechanical properties* of the main $LaFe_{11.2}Si_{1.8}$ 1:13 phase and its minority phases α-Fe and $La_1Fe_1Si_1$ were assessed by Vickers microhardness and nanoindentation testing. The microhardness was measured with a pyramidal-shaped Vickers diamond indenter and a load of 300 gf (HV0.3) using a Buehler Micromet 5104 machine. The microhardness was measured only for the 1:13 phase since the size of the Vickers impressions was comparable with the sizes of the largest secondary phase grains. Over 30 Vickers indents separated by a distance of more than 30 times the indent diagonal were evaluated. Depth-sensing nanoindentation was carried out on a platform Nanoindenter (Keysight Tec, Inc.) equipped with a three-sided pyramidal Berkovich diamond tip (Synton, Switzerland) and a continuous stiffness unit in order to measure the contact stiffness continuously over indentation depth. The tip area functions and the machine stiffness were calibrated using fused silica as a reference material and the local mechanical properties in terms of hardness and modulus of all three phases were calculated according to the Oliver-Pharr method [27]. The Poisson's ratio was assumed as 0.29 for the α-Fe phase and as 0.28 for the 1:13 and $La_1Fe_1Si_1$ phases. All nanoindentation experiments were performed in constant strain-rate mode by applying a preset indentation strain-rate of 0.05 $s^{-1}$ to an indentation depth of 100 and 1000 nm, and at least eight individual indentations per phase were taken. Neighboring indents were separated by a distance of 20 times the indentation depth.

## 3. Results and discussion

### 3.1 Magnetocaloric properties and microstructure

The transition temperature of the magnetocaloric $LaFe_{11.2}Si_{1.8}$ material was estimated by the first derivative of the magnetization versus temperature curve $M(T)$ and is about $T_t = 214$ K with marginal thermal hysteresis width (c.f. Fig. 1, inset). The magnetic entropy change reaches a value of about $\Delta S_{mag} = -9.6$ $Jkg^{-1}K^{-1}$ (c.f. Fig. 1). The measured $\Delta S_{mag}$ value for this alloy composition is comparable to results reported earlier [28, 29]. The microstructure of the material tested in this study is presented on the SEM micrograph shown in Fig. 2. The samples include a small amount of the minority phases α-Fe and $La_1Fe_1Si_1$ (also called below 'La-rich phase') which appear as black and white areas on the image, respectively. Additional XRD (not shown) revealed that the amount of minority phases is about 5 wt.%. The grain sizes of the $LaFe_{11.2}Si_{1.8}$ main phase are distributed in the range between 5 μm and 15 μm. Both secondary phase inclusions (α-Fe and $La_1Fe_1Si_1$) have typical sizes between 1 μm and 20 μm equivalent diameter.

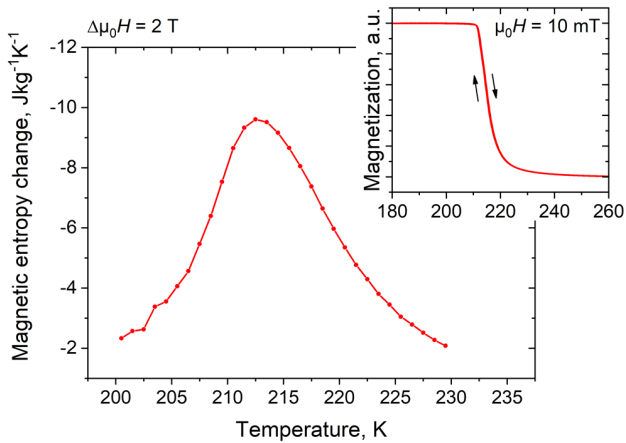

Fig. 1. Magnetic entropy change $\Delta S_{mag}$ vs. temperature $T$ of a $LaFe_{11.2}Si_{1.8}$ sample for a magnetic field change $\Delta \mu_0 H = 2$ T. The inset shows the corresponding magnetization $M$ vs. temperature $T$ plot ($\mu_0 H = 10$ mT).

Fig. 2. Typical SEM micrograph (BSE contrast) of the microstructure observed in the LaFe$_{11.2}$Si$_{1.8}$ samples. One can clearly distinguish the grain structure of the main phase (different shades of the grey matrix) and the secondary phases α-Fe (black) and La-rich (white).

### 3.2 Global mechanical properties

#### 3.2.1 Fracture stress, strain and Young's modulus obtained from compression tests

During compression testing of LaFe$_{11.2}$Si$_{1.8}$ catastrophic brittle failure was observed in most cases meaning that the sample fractured in an explosion-like manner, splitting into dozens of small pieces. In some cases it was possible to observe formation of cracks or splitting of the specimen parts prior to total fracture. An example of such a sample is shown in Fig. 3 and the corresponding video of the whole compression test is given in the Supplementary video V1. Fig. 3a displays the sample surface (depicted by the dashed rectangle) before the compressive force is applied. In Fig. 3b one can see the same sample with numerous cracks formed during compression testing prior to total fracture. In order to guide the eye, the cracks are tracked more distinctly in the schematic illustration depicted in Fig. 3c. By analyzing the orientations of the cracks with respect to the loading direction one can suggest that there are two main fracture mechanisms involved. The first one is pure cleavage fracture due to the transverse tensile stress, which is manifested by a crack running parallel to the applied force. This is the main fracture mechanism, which was observed in all cases when the fracture was less catastrophic, without an explosion-like fragmentation. The second mechanism is shear fracture, which is manifested by the cracks running at an angle close to 45° with respect to the loading direction (Fig. 3c). The fracture morphology observed in LaFe$_{11.2}$Si$_{1.8}$ samples is very similar to the behavior of rock samples or ceramics under uniaxial compression. Both tensile cleavage and shear cracking are observed in marble [30] cement mortar [31, 32], granite [33], sandstone [34] and technical ceramics [35].

Fig. 3. Characteristic photographic images of LaFe$_{11.2}$Si$_{1.8}$ sample during the compression test. (a) corresponds to unstrained sample, (b) shows the same sample prior to total fracture, and (c) is a schematic illustration showing the two types of cracks which are observed in (b). The compression direction is vertical.

The images of the sample surface during a compression test were utilized for the measurements of the sample strain by digital image correlation (DIC). In Fig. 4a the surface of a sample with the spray pattern is shown along with the positions of four digital extensometers depicted by the double-headed arrows, and the corresponding stress-strain curves are shown in Fig. 4b. As one can see, all four extensometers generally follow the same linear trend. Small jumps in the extensometer data are attributed to the uncertainties of the digital image correlation caused on one hand by small changes of the surface pattern contrast during loading and on the other hand by very small measured displacements of only few micrometers, which are at the limit of the resolution of the camera. An important requirement for DIC measurements is an unchanged sample surface. If the surface contrast changes (e.g. due to formation of cracks) then proper image correlation is not possible or will bring unreliable results. Thus, the strain values were recorded only if no significant changes of the surface contrast occurred. The fracture of the samples was defined by the moment when the applied force dropped to zero.

The averaged Young's modulus out of 15 samples obtained from the current experimental setup was calculated to be $E_{avg} = 97\pm23$ GPa. This value corresponds well to the modulus obtained from resonance frequency and damping analysis of a La(Fe, Co, Si)$_{13}$ alloy by Kaeswurm et al. [18]. The mechanical strain measured by DIC scatters from sample to sample and even between different areas on the surface of the same sample. It was also observed that small irregularities as well as non-plane-parallelism of the top and bottom surfaces can lead to an inhomogeneous distribution of the surface strain. Most probably the same factors were responsible for the very large spread of the measured strains of a La(Fe, Co, Si)$_{13}$ alloy reported by Löwe et al. [21], where the fracture strains are between $\varepsilon = 0.05$ % and 0.25 %. In the present study fracture of the LaFe$_{11.2}$Si$_{1.8}$ samples was observed at strains between $\varepsilon = 0.2$ % and 0.6 % and stresses between $\sigma = 180$ to 620 MPa. Due to this large spread of the measured data the fracture properties of LaFe$_{11.2}$Si$_{1.8}$ cannot be properly described by a single fracture stress or strain value. Instead, a statistical analysis, as it is typically implemented for the characterization of brittle ceramic materials, is employed in the next subsection.

Fig. 4. Digital image correlation measurements of the sample strain. The image of a sample with four digital extensometers is depicted in (a). The corresponding engineering stress-strain curves for the four digital extensometers are shown in (b).

3.2.2 Statistical description of compressive failure

Due to the probabilistic character of the measured critical stresses, the fracture of brittle materials is typically characterized by a Weibull distribution [36]. In the field of ceramics, the Weibull distribution of the fracture behavior is explained by the concept of the 'weakest link'. Weak links are considered to be micro-structural features, such as cracks, pores, minority phases, grain and phase boundaries or surface notches. Such imperfections govern the mechanical properties, since the mechanical stress concentrated around them cannot be compensated by the motion of dislocations in brittle materials. The cumulative probability of fracture, $P_f$, is given by Eq. 2

$$P_f = 1 - exp\left(-\left(\frac{\sigma}{\sigma_0}\right)^m\right), \qquad (2)$$

where $\sigma$ is the measured fracture stress, $\sigma_0$ is the characteristic strength corresponding to the stress at which 63.2 % of samples fail, and $m$ is the Weibull modulus describing how reliable is the prediction of the material failure. A large $m$ indicates a reliable prediction of the failure behavior of the material.

It is well-known that compressive fracture of brittle specimens depends on the sample size [37, 38, 39]. In order to estimate the effect of the sample size two sets of data are plotted in a Weibull plot, as shown in Fig. 5. The black squares depict the samples with the aspect ratio of approximately 1:1.2 while the red triangles correspond to the samples with an aspect ratio of approximately 1:2.5.

Fig. 5. Weibull distribution of the fracture stress for LaFe$_{11.2}$Si$_{1.8}$ samples with two different aspect ratios. The characteristic stress $\sigma_0$ is 403 MPa for the aspect ratio of 1:2.5 and 560 MPa for the aspect ratio of 1:1.2. The dashed circle marks the early failures.

There are several distinct features on the Weibull plot which need to be discussed. First of all, there is an increased probability of failure at low stresses as depicted by the dashed circle. We attribute these early failures to sample misalignment or non-plane-parallel top and bottom faces. Secondly, the fracture stresses measured for the samples with an aspect ratio of 1:1.2 are significantly higher in comparison to the values obtained for an aspect ratio of 1:2.5. The values of the characteristic strength $\sigma_0$ are calculated to be about 403 MPa for the aspect ratio of 1:2.5 and 560 MPa for the aspect ratio of 1:1.2. Such a significant difference in the characteristic strength values points out the importance of testing conditions and sample geometry for the evaluation of the materials properties and can partially explain the difference in the measured mechanical properties of La(Fe, Si)$_{13}$ reported by different groups. The effect of sample size on brittle fracture is typically explained by the statistical distribution of critical flaws since the probability of having a critical flaw increases with increasing sample volume.

Due to the non-linearity of the distributions caused by early failures it is challenging to extract an unambiguous Weibull modulus from Fig. 5. Depending on how the linear fit is conducted, the values of $m$ vary between 2 and 6, which reflects that the prediction of the material failure for the current material, sample preparation and testing method is rather uncertain. A Weibull modulus of $m = 0$ to 10 is characteristic for ceramics like SiC or Al$_2$O$_3$ [35]. The analogy between ceramics and the intermetallic phase La(Fe, Si)$_{13}$ is valid, as both material classes include strong covalent bonding between the atoms, but intermetallic phases also include the character of conventional metallic bonding. The covalent bonding in La(Fe, Si)$_{13}$ is in between Fe and Si [40, 41]. As the NaZn$_{13}$-type unit cell of La(Fe, Si)$_{13}$ with a lattice parameter depending on the Si content from $a = 1.1487$ to $1.1439$ nm ($T = 300$ K) [42] is large and involves over 100 atoms with huge difference in atomic diameter (up to 35 % between Si and La), there are presumably no mobile dislocations in this highly ordered class of material, which explains its brittleness and lack of deformability. Since no single-phase material was obtained up to now in La(Fe, Si)$_{13}$-based alloys, the intrinsic properties of the La(Fe, Si)$_{13}$ phase can only be assessed by local mechanical testing methods, e. g. the nanoindentation method as described in the following subsection.

Table 1: Summary of global and local mechanical properties of the LaFe$_{11.2}$Si$_{1.8}$ alloy and its secondary phases.

| Mechanical property | | Global | |
|---|---|---|---|
| Fracture stress, compression test | | 180 – 620 MPa | |
| Fracture strain, compression test | | 0.2 – 0.6 % | |
| Young's modulus, compression test | | 97±23 GPa | |
| **Mechanical property** | **1:13 phase** | **α-Fe phase** | **La-rich phase** |

| Hardness, Vickers | 680±40 HV0.3 | | |
|---|---|---|---|
| | ~6.7±0.4 GPa | | |
| Hardness, nanoindentation: indentation depth 1000 nm | ~10 GPa | | |
| Hardness, nanoindentation: indentation depth 100 nm | 14.8±0.6 GPa | 3.4±0.3 GPa | 12.5±0.6 GPa |
| Local Young's modulus, nanoindentation | 173±6 GPa | 200±3 GPa | 155±3.4 GPa |

### 3.3 Characterization of the local mechanical properties of LaFe$_{11.2}$Si$_{1.8}$ by indentation

For the sake of clarity the local mechanical properties measured by indentation techniques described in this subsection are summarized in Table 1 together with the global mechanical properties discussed in the previous subsections.

The details of a typical nanoindentation experiment for three phases of LaFe$_{11.2}$Si$_{1.8}$ are displayed in Fig. 6. The load-displacement data (Fig. 6a) represent a kind of fingerprint for the tested phases, from which the potential of elastic-plastic deformation of each component can be concluded. While the main 1:13 phase and the La-rich phase exhibit a quite large elastic unloading component resulting in a residual indentation depth of only 50-60 nm, the indentation in the α-Fe phase leads to fully plastic deformation with a residual imprint of 100 nm indentation depth. Moreover a pronounced pop-in behavior for the main phase at ~30 nm can be clearly seen (depicted by a grey oval in Fig. 6a). Such pop-in events were observed during all indentations of the main phase and correspond to the initial yield point of the material.

Fig. 6. Results of a typical nanoindentation experiment for the main phase and the secondary phases of LaFe$_{11.2}$Si$_{1.8}$. Load-displacement curves for the three phases are shown in (a). The corresponding Young's modulus and hardness continuously recorded over indentation depth are shown in (b) and (c), respectively.

The corresponding continuously monitored Young's moduli, displayed in Fig. 6b, exhibit a rather weak dependence over indentation depth, with some minor scatter due to the experimental determination of the contact stiffness. The fact that the Young's moduli are nearly constant with increasing indentation depth proves that the nanoindentation setup was properly calibrated. The Young's moduli of all three phases determined by the nanoindentation method do not differ significantly and range between 155 and 200 GPa. One can notice that there is also a significant difference between the Young's modulus of the main phase measured by nanoindentation (173±6 GPa) and the global value obtained from the compression tests (97±23 GPa). Although compression testing is not a preferable method to measure Young's modulus one should notice that a similar value of about 100 GPa was obtained by resonance frequency and damping analysis (RFDA) in [18]. The lower values of the Young's moduli can be explained by the fact that both RFDA and compressive tests are applied to much larger sample volumes containing some porosity, cracks and numerous phase boundaries while nanoindentation probes essentially local intrinsic properties of the material. One should also notice that the Young's modulus measured by nanoindentation in the present work (173±6 GPa) is significantly higher than the Young's modulus of La(Fe, Co, Si)$_{13}$ measured in [18] (130±5 GPa). It is currently unclear whether this difference is caused by addition of Co or by other factors.

Continuously monitored hardness values up to indentation depths of 100 nm are presented in Fig. 6c. The main phase and the La-rich have relatively high hardness (14.8±0.6 GPa and 12.5±0.6 GPa, respectively) comparable with ceramic materials. The hardness of the α-Fe phase (3.4±0.3 GPa) is roughly two times larger than the hardness of pure iron [43] but corresponds well to the hardness of the ferrite phase in steels [44, 45] indicating that the α-Fe grains contain some amount of Si. The main phase also demonstrates a distinct feature corresponding to the pop-in event in the load displacement curve. While under 30 nm the hardness is increasing linearly in the elastic regime, after the pop-in the value is nearly constant. Although no clear indentation size effect is observed for indentation depths up to 100 nm, additional deeper indentations in the main phase with a preset depth of 1000 nm showed some size effect leading to a lower measured hardness of about 10 GPa. The Vickers microhardness expressed in SI units [46] was measured to be about 6.7±0.4 GPa. Taking into account that Vickers indenters have even higher indentation depths up to 10 μm, one can attribute the differences in the measured hardness values to a size effect meaning that the measured hardness increases with decreasing the indentation size [47, 48]. An illustrative comparison of different indents in LaFe$_{11.2}$Si$_{1.8}$ is given in the Supplementary material, Fig. S1. It should be however kept in mind that a direct comparison of nanoindentation hardness and Vickers hardness is not straightforward due to differences in the tip geometry, different specifications on the hardness analysis as well as different plastically deformed volumes [49, 50].

In order to correlate the results of indentation and compressive testing the widely used relationship between the strength and hardness can be utilized [51, 52]:

$$H_V \sim C_F \cdot \sigma_y, \qquad (3)$$

where $C_F$ is the constraint factor, $H_V$ is the Vickers hardness and $\sigma_y$ is the yield strength. For metallic materials with elastic-plastic behavior, the constraint factor is about 3 [51, 52]. For ceramics which exhibit only very restricted plasticity it can be even lower [53]. Taking the value of $C_F = 3$ and using the measured Vickers hardness of ~6.7±0.4 GPa, the lower bound of the strength of the main phase is estimated to be 2 GPa, which is four times higher than the strength values obtained from the compression tests in this work and an order of magnitude higher than the values reported elsewhere [18, 21]. Even though Eq. 3 can be used only as a rough estimate, one can conclude that the compression strength of macroscopic samples does not reflect the actual material property but is rather determined by the material defects. With decreasing sample size, or, in the case of indentation, the material volume subjected to mechanical load, the detrimental effect of defects becomes less pronounced.

**3.4 The role of secondary phases and intrinsic defects**

As was mentioned above, the brittle fracture behavior is usually determined by the material defects or flaws, which act as crack initiators. Apart from micropores or microcracks, which always exist in a bulk material, also the role of secondary phases in LaFe$_{11.2}$Si$_{1.8}$ in crack initiation and fracture needs to be clarified. Due to the large volume of samples used in compression experiments it seems to be virtually impossible to clearly specify which kind of defect or flaw is responsible for the initiation of cracks. Instead, we suggest to consider the interaction of indentation-induced cracks with secondary phases, grain boundaries and phase boundaries. Fig. 7 presents an example of such an analysis. The SEM image (Fig. 7a) shows the area containing several α-Fe grains. The EBSD grain orientation map of the same area is shown in Fig. 7b. To induce cracks, a Vickers indent was made in the vicinity of the α-Fe grains, as shown in Figs. 7c-d. The crack propagates from the corner of the indent and is stopped by the α-Fe grain. Iron is known to be a ductile material, which can deform plastically to large extent without cracking [54]. According to this, when the crack approaches an α-Fe grain, it deforms plastically dissipating the energy required for crack propagation. More examples of crack arresting by α-Fe grains can be found in the supplementary material, Fig. S2. The arrest of a propagating crack on the interface between dissimilar materials is a well-known effect and is described in detail, for instance, in [55, 56] for the case of alternating soft and hard aluminum layers. A possible way to improve the mechanical stability of La(Fe, Si)$_{13}$-type alloys would be to take advantage of the crack arresting action of α-Fe grains. A proper network of thin α-Fe inclusions might have a vital effect on the fracture strength of macroscopic samples. However, a recent paper [57] where this idea was realized by adding excess Fe into a La(Fe, Si)$_{13}$ alloy demonstrated only a minor increase of the measured strength with a simultaneous decrease of the magnetocaloric effect due to the extra iron volume. Another important conclusion which can be drawn from Fig. 7 is that the cracks can easily propagate both through the grains of the main phase and along the grain boundaries. The main crack starts to propagate from the corner of the indent along the grain boundary but then it runs through a grain before reaching the α-Fe grain. There is also a small crack marked by white arrows in Figs. 7c and 7d, which is clearly running through the grains. In other words, there is no preferable propagation path for indentation-induced cracks within the main phase.

In contrast to the α-Fe grains, the La-rich grains play a detrimental role for the mechanical stability of La(Fe, Si)$_{13}$. As can be seen in Fig. 2, virtually all La-rich phase grains have pre-existing cracks. Detailed examination of the inset in Fig. 2 reveals that even the smallest La-rich grains exhibit a crack. A FIB cut shown in the Supplementary Fig. S3 demonstrates that these cracks are not restricted to the surface area but are extended in the bulk. Taking into account the brittleness of the whole alloy such pre-existing cracks definitely represent potential weak links for the initiation of catastrophic fracture. Preferable cracking around the La-rich grains was directly observed during cyclic magnetocaloric transition in [9]. The nature of these pre-existing cracks is unclear. According to the nanoindentation data (Table 1) there is no dramatic difference in Young's modulus or hardness between the main phase and the La-rich phase. Such kind of cracks would be expected to occur during cooling from 1050 K to room temperature if there is a significant difference in the coefficient of thermal expansion between the La-rich and the 1:13 phase.

Fig. 7. Interaction of surface cracks with the microstructure of LaFe$_{11.2}$Si$_{1.8}$. The SEM micrograph and the corresponding EBSD grain orientation map of the original surface are shown in (a) and (b), respectively. The same surface area after the Vickers indent is made is depicted in (c) and (d), respectively. The white arrows in (c) and (d) point onto a crack which is running purely through the grains.

## 3.5 Discussion of mechanical properties in relationship with the magneto-volume transition in La(Fe, Si)$_{13}$

The link between the investigated mechanical properties and the mechanical stability of La(Fe, Si)$_{13}$ during magnetocaloric transitions, e. g. during magnetic field cycling or thermally induced transitions, is briefly discussed in this section. The magnetocaloric transition of La(Fe, Si)$_{13}$ is accompanied by a volume change that induce stresses $\sigma_V$, which can be *roughly* estimated via the bulk modulus $K$ according to [35, 58] via equation 4:

$$\sigma_V = K\varepsilon_V = \frac{E}{3(1-2v)}\left(\frac{\Delta V}{V}\right), \qquad (4)$$

where $E$ is the Young's modulus, $v$ is the Poisson's ratio, and $\Delta V/V$ is the relative volume change. To calculate $\sigma_V$, the Young's modulus $E$ of the 1:13 phase obtained in this work, $v = 0.3$ as an estimate and $\Delta V$ from [29] are used below. Although the Young's modulus was determined at ambient conditions, it is known [18], the modulus decreases by only 10-15 % with decreasing temperature down to the phase transition. Thus, for estimation purposes we assume the Young's modulus between 140 and 170 GPa. It is known that the volume change during the magnetocaloric transition of LaFe$_{13-x}$Si$_x$ increases with decreasing Si content: for $x = 1.2$ the volume change is $\Delta V_{x=1.2} = 1.5$ vol.%, for $x = 1.6$ the volume change is $\Delta V_{x=1.6} = 1.0$ vol.%, and for $x = 1.8$ the volume change is $\Delta V_{x=1.8} = 0.8$ vol.% [29], respectively. Thus, according to Eq. 4 the volumetric stresses should increase with decreasing Si content: $\sigma_{V\,x=1.2} = 2.2$ GPa, $\sigma_{V\,x=1.6} = 1.5$ GPa and $\sigma_{V\,x=1.8} = 1.1$ GPa for E = 170 GPa while $\sigma_{V\,x=1.2} = 1.8$ GPa, $\sigma_{V\,x=1.6} = 1.2$ GPa and $\sigma_{V\,x=1.8} = 0.9$ GPa for E = 140 GPa. This estimate indicates that all LaFe$_{13-x}$Si$_x$ alloys with $x = 1.2$ to 1.8 should suffer from stresses exceeding the fracture stresses measured by compression testing in this work or elsewhere [18, 21]. However, pure cycling of LaFe$_{13-x}$Si$_x$ with $x = 1.8$ in a magnetic field or even multiple thermally induced transitions with a liquid nitrogen bath does not create any cracks (not shown). This fact supports our conclusion that compressive fracture is not a proper indicator of the intrinsic strength of the main phase. In contrast, the estimation of the strength of the 1:13 phase of 2 GPa for $x = 1.8$ made on the basis of the indentation results (Eq. 3) seems to be correct, as it is well above the critical stress of $\sigma_{V\,x=1.8} = 0.9$ to 1.1 GPa calculated from Eq. 4. Cycling of samples with low Si content ($x = 1.2$) induces volumetric stresses of $\sigma_{V\,x=1.2} = 1.8$ to 2.2 GPa leading to a fragmentation of the sample, which is shown for a thermally cycled sample in Fig. 8 [10], in good correspondence with the estimation of Eq. 3.

Fig 8: X-ray computed tomography images with 3 μm spatial resolution of thermally cycled LaFe$_{13-x}$Si$_x$, x = 1.2. White arrows are pointing on cracks. Thermal cycles were performed via sequential temperature sweeping around the transition temperature in a PPMS-VSM magnetometer ($N = 50$ cycles, $\dot{T} = 2$ Kmin$^{-1}$), from [10].

## 4. Conclusions

Investigations of the mechanical properties of LaFe$_{11.2}$Si$_{1.8}$ and its minority phases at different length scales are presented. The Young's modulus estimated from the compression tests of macroscopic samples is $E_{global} = 97\pm23$ GPa while the local modulus of the 1:13 phase of $E_{local} = 173\pm6$ GPa is measured by nanoindentation. The fracture of $LaFe_{11.2}Si_{1.8}$ during a compression test occurs at strains below $\varepsilon = 0.6\%$ and is brittle in nature, similar to the fracture of rock or ceramic samples. Statistical analysis of compression tests reveals a strong dependence of the measured compressive strength on the geometry of the sample. Samples with the same cross-sectional area but smaller aspect ratio can sustain larger stresses ($\sigma_0 = 560$ MPa) in comparison to the samples with larger aspect ratio ($\sigma_0 = 403$ MPa). This difference clearly indicates that compressive fracture is determined by the statistics of pre-existing defects such as pores, cracks and phase boundaries. Also nanoindentation reveals some size effect since the hardness obtained with 100 nm indentation depth is about 15 GPa but for the indentation depth of 1000 nm a hardness of 10 GPa is measured. A hardness of 6.7 GPa is measured by Vickers microhardness tests. Estimated from the indentation results, the magnetocaloric 1:13 phase should possess a strength of about 2 GPa, which is significantly larger than the fracture stress obtained from compressive tests and is sufficient to survive cyclic phase transitions required during operation of a magnetocaloric devices. Thus, in order to access intrinsic mechanical properties of the magnetocaloric phase, indentation-based methods are preferable over global mechanical testing methods like compression or bending tests.

Investigations of the interaction of indentation-induced cracks with the microstructure of the $LaFe_{11.2}Si_{1.8}$ alloy reveals that the cracks propagate both along the grain boundaries as well as through the grains of the main phase. Propagating cracks can be efficiently stopped by α-Fe grains. On the contrary, the grains of the La-rich secondary phase virtually always contain pre-existing cracks providing numerous sites for fracture initiation.

To outline further research directions one should note, that important information about the basic mechanical properties of $La(Fe, Si)_{13}$ alloys is still missing. For instance, the *fracture toughness* – the main parameter describing how easy a crack can propagate – is not known for the main phase. In our opinion, micromechanic experiments are required in order to access the intrinsic mechanical properties of $La(Fe, Si)_{13}$ alloys in more details. In a micromechanic experiment, micron-sized samples are prepared and subjected to mechanical loading in-situ inside a scanning electron microscope. In this way the stress-strain curves, strength, and fracture toughness of single crystal specimens with different orientations as well as of samples containing specific grain or phase boundaries can be measured. Also, temperature dependent experiments should be performed to investigate the mechanical behavior at the transition temperature.

Acknowledgments

O.G. acknowledges support by the Austrian Science Fund (FWF), project P27432-N20. Financial support by the Austrian Federal Government (837900) within the framework of the COMET Funding Programme (MPPE, project, A7.19) is appreciated (V.M.-K.). This work has been financially supported by the German Research Foundation in the framework of SPP1599 "Ferroic Cooling" under grant WA3294/3-2 (A.W.) and by the Germany Federal Ministry for Economic Affairs and Energy under the project number 03ET1374B (M.K.).